# Persistent coherence of quantum superpositions in an optimally doped cuprate revealed by 2D spectroscopy

Fabio Novelli[1], Jonathan O. Tollerud[1], Dharmalingam Prabhakaran[2], Jeffrey A. Davis[1,3,*]

[1] Centre for Quantum and Optical Science, Swinburne University of Technology, Hawthorn, Victoria 3122, Australia

[2] Department of Physics, University of Oxford, Oxford OX1 3PU, UK

[3] ARC Centre of Excellence for Future Low-Energy Electronics Technologies, Swinburne University of Technology, Hawthorn, Victoria 3122, Australia

[*] jdavis@swin.edu.au

*Understanding of the precise mechanisms of high-temperature superconductivity is elusive. In particular, in order to solve the puzzle of the pairing mechanism, it is important to understand the detailed nature of the excitations at energies around the superconducting gap. While measurements of the dynamics of excited electronic populations have been able to give some insight, they have largely neglected the intricate dynamics of quantum coherence. Here, we apply multidimensional coherent spectroscopy for the first time to a prototypical cuprate and report unprecedented coherent dynamics persisting for ~500 fs, originating directly from the quantum superposition of optically excited states separated by 20 – 60 meV. These results reveal the correlation between high and low energy excitations, and indicate that the interplay between many-body states on different energy scales conserves phase coherence. In revealing these dynamics we demonstrate that multidimensional coherent spectroscopy can address electronic correlations and interrogate many-body quantum systems in unprecedented ways.*

**Introduction**

In conventional superconductivity, electronic excitations are "glued" into Cooper pairs by lattice vibrations(*1, 2*). It has been suggested that a different kind of glue, involving for example excitations in the mid-infrared (MIR), could be responsible for cuprate superconductivity(*3, 4*). However, understanding the nature of these excitations and the absorption spectrum of the cuprates is puzzling even at equilibrium(*5*). For undoped systems, which appear as shiny metals to the naked eye, the optical absorption is dominated by charge-transfer gap transitions around 2 eV with a broad, asymmetrical and long absorption tail which extends to lower frequencies. When the system is chemically doped, the absorption below ~2 eV increases at the expense of spectral weight above the gap(*6*): the material develops a Drude peak with a frequency-dependent scattering rate(*5*), and a broad continuum of MIR excitations with different symmetries(*7*). Both model calculations(*8–10*) and the similarly broad features observed in Raman spectroscopy(*11, 12*) suggest that this broad continuum involves, in addition to magnetic(*13–15*) and vibrational(*5, 6*) transitions, many body excitations of different, mixed atomic orbitals.

In order to address the interactions that lead to superconductivity in cuprates, a variety of pump-probe time-resolved experiments have been performed(*16–22*), which allow mapping of the dynamics of excited electronic populations and measurement of the electron-boson couplings(*23–27*). Oscillations in pump-probe traces are indicative of a coherent response, and there are many examples in the literature of such oscillations in strongly-correlated electronic materials(*28–30*). However, in a typical pump-probe experiment, it is difficult to distinguish between competing signal pathways and identify the origin of the coherent response. In most cases the oscillations in pump-probe data arise from displacive excitation of coherent phonons(*31–33*), usually optical phonons. In this case, an initial electronic excitation and strong electron-phonon coupling drive the excitation of the coherent phonons(*31*). Coherent acoustic phonons, excited by a stress gradient at the sample surface due to the laser absorption, have also been shown to generate oscillations as a result of the probe beam reflecting off the front of the shock wave and interfering with the reflection off the sample surface(*34*). Mansart *et al*. have reported observations of coherent oscillations in pump-probe measurements which they attributed to oscillations of the Cooper pair condensate(*30*). To make that attribution, the relevant oscillations were extracted from a series of pump-probe measurements and a list of assumptions are made about the excitation process being impulsive stimulated Raman scattering. While potentially opening the door to new physics, this highlights one of the limitations of pump-probe measurements for revealing coherent dynamics: many different pathways contribute to the non-linear signal and, while coherent Raman-like pathways are expected to be present, the signal tends to be dominated by population effects and it is difficult to disentangle the different contributions.

Pump-probe experiments are a subclass of transient four-wave mixing (FWM) techniques that have been used for decades to measure dynamics(*35, 36*). In a FWM experiment on a simple two-level system, the first pulse generates a coherent superposition between ground and excited state and the interaction with the second pulse converts this into a population. (In pump-probe measurements, these first two interactions come from a single pulse: the pump). The third pulse converts this population into a coherent quantum superposition, which relaxes back to the ground state emitting the signal. (In pump-probe measurements the third pulse is the probe, the signal is emitted in the same direction as the probe, and the interference between them gives the change in probe intensity). For more complex samples which cannot be described as two-level systems, other pathways become possible, for example, instead of the first two pulses combining to excite a population, they can excite a coherent superposition of closely spaced energy states, in much the same manner as a coherent Raman experiment(*37*). Single-colour transient grating experiments are another subclass of FWM, which has been applied to reveal the quasi-particle diffusion coefficients and scattering rates in YBCO(*38*). Like pump-probe experiments, these measurements probe population dynamics.

Other FWM approaches, however, can probe directly the coherent dynamics of quantum superpositions(*39–42*). To our knowledge, the measurement of such dynamics in a strongly correlated many-body electronic system has not been investigated previously. Multidimensional coherent spectroscopy (MDCS)(*43*) is an approach to FWM capable of precisely measuring coherent dynamics. It has been widely applied to semiconductors(*44*), molecules(*45*), and proteins(*46–48*), where it has enabled detailed understanding of relevant interactions. The key enhancement in MDCS over standard FWM experiments is that the phase of the signal is measured in addition to the amplitude. This is achieved by ensuring the three excitation pulses are phase-locked with each other and with a fourth pulse, the local oscillator, which is overlapped with the signal for heterodyne detection. By recording the phase of the signal, both the real and imaginary parts of the response are simultaneously measured without the need of Kramers-Kroenig transformation.

Through the direct measurement of the phase information in MDCS it becomes simpler to disentangle different pathways, allowing a greater level of control and insight into the interactions in strongly-correlated electronic materials. Pathway-selective MDCS provides additional capability to separate different contributions: by controlling the spectral phase and amplitude of the excitation pulses, one can trigger and probe specific components or pathways of the non-linear response of a material, such as pure quantum-coherent signals(*49*). Here we use this approach to selectively excite and probe the dynamics of coherent quantum superpositions of low-energy excitations in a typical cuprate superconductor (LSCO, x=15%, $T_C$~38 K). We found coherent dynamics persisting for ~500 fs originating directly from the quantum superposition of optically-driven many-body states with energy separation of 20 – 60 meV. Interactions with the third optical pulse reveal the correlation between high and low energy excitations.

**Results and Discussion**

**Experimental description.** The experiment is described schematically in Fig.1: the first pulse, with wave-vector $k_1$ is spectrally shaped to be centred at 1.575 eV with full-width at half maximum (FWHM) of 20 meV; the second and third pulses, with wave-vector $k_2$ and $k_3$, respectively, are centred at 1.54 eV with FWHM 30 meV, such that there is minimal spectral overlap with the first pulse (see Fig.1a for the experimental geometry and Fig.1b for the spectra of the pulses). The first two pulses, which are set to arrive simultaneously, excite a coherent superposition between states, as indicated by the black oscillating curve in Fig.1d and Fig.1e. The spectral width of the pulses means that only quantum superpositions between states with energy separations ranging from about 20 meV to 60 meV will contribute to the measured signal beyond pulse overlap (Fig.1c). There are many possible pathways by which these coherences could be excited, based on where the optical pulses are being absorbed/emitted to/from (see Fig.S1 in the SI). Two such signal pathways are shown in Fig.1e.

For the purposes of our discussion here we focus on the state of the system after the first two pulses, at which point, the sample is in a coherent quantum superposition of states separated by 20 – 60 meV. The third pulse, with energy 1.54 eV and wave-vector $k_3$, then interacts with this coherent superposition, leading to the 3$^{rd}$ order signal emitted in the phase-matched direction $k_s=-k_1+k_2+k_3$, with central energy expected at ~1.505 eV. The emitted signal is overlapped with the local oscillator and sent to a spectrometer where a spectral interferogram is measured and analysed to provide the amplitude and phase of the signal. To probe the evolution of the coherent superposition excited by the first two pulses, the delay between the second and third pulse, $t_2$, is scanned and the evolution of the signal amplitude and phase is measured. The sample was at a nominal temperature of 12 K but quantitatively consistent results were also obtained at 66 K (Fig.S2). This indicates that we are not sensitive to the formation of the superconducting phase, which is melted because of the high repetition rate of 93 MHz (see SI for details).

**Measurements results.** The real part of the measured signal is shown in Fig.2a as a function of emission energy and delay time $t_2$; the integrated amplitude of the signal as a function of $t_2$ is shown in Fig.2b. These clearly show the signal persisting out to $t_2 \sim 500$ fs, indicating that the quantum superpositions excited by the first two pulses remain coherent over this timescale. Because we measure the phase of our signal, the 2D plot in Fig.2a shows oscillations as a function of $t_2$, corresponding to the phase of the coherent superpositions that are excited in LSCO. Intriguingly, the frequency of these oscillations appears to change as a function of the emission energy. To gain more insight into this signal, the data is Fourier transformed with respect to $t_2$ to generate the 2D spectrum in Fig.2d. This 2D spectrum shows a narrow peak elongated along the 1:1 diagonal on top of a broader signal due to the response at pulse overlap. The cross-section along the anti-diagonal, shown in the inset of Fig.2d, reveals a triangular peak shape typical of a Lorentzian on top of a broader Gaussian profile. By windowing the data in the $t_2$ domain we can separate the response during pulse overlap, which contributes the broad part of the 2D spectrum, from the response after pulse overlap, as shown in Fig.2e and Fig.2f, respectively.

The 2D spectrum corresponding to the signal persisting beyond pulse overlap (Fig.2f, obtained by windowing the data with the red function displayed in Fig.2c) indicates that a signal is generated for a large distribution of coherence energies ($E_2$) or, in other words, that a broad distribution of low-energy states are coherently excited and contribute to the extended signal. The projection of the data onto the $E_2$ axis matches closely the distribution of energy differences expected between the first two pulses, as shown in Fig.1c. For such a large distribution of coherence energies, one would normally expect the macroscopic coherence of the ensemble, and hence the measured signal, to decay rapidly (<50 fs) due to the different frequency components evolving to be out of phase. The expected 2D spectrum would then appear as a broad peak, matching the contribution that arises during pulse overlap (Fig.2e). However, the results show that right across the distribution of coherence energies there is a component that remains coherent for over ~500 fs, generating the narrow peak elongated along the diagonal in Fig.2d and Fig.2f, with a cross-diagonal width of ~9 meV.

The narrow diagonal peak is reminiscent of what is obtained in photon echo measurements, where rephasing of different frequency allows homogeneous linewidths to be measured even in the presence of significant inhomogeneous broadening(*39*, *50*). In the present case, however, there is no rephasing, and the ability to observe the coherent evolution of quantum superpositions with a broad distribution of coherence energies is possible only because there is a correlation between the coherence energy ($E_2$) and the emission energy ($E_3$). This correlation is also evident in Fig.2a, where the frequency of the oscillations varies as a function of the emission energy. This is mapped onto the 2D spectrum, and leads to the narrow diagonal peak in Fig.2f. The cross-diagonal width of this peak is then determined by the strength of the correlation, and/or by the decoherence of the quantum superpositions.

**Comparison with vibrational coherence.** 2D spectra obtained for samples with clear discrete vibrational modes, as is the case for coherent optical phonons, show narrow peaks in $E_2$ which are spread across a range of $E_3$ values, as shown in Fig.3 for a laser dye molecule. This is typical of the 2D spectra normally expected for this type of experiment, and is in stark contrast to what we observed for LSCO. Based on the time evolution of the signal amplitude for LSCO (Fig.2b), and in the absence of the phase information, one would expect a narrow peak along the $E_2$ axis, similar to those observed for the laser dye (Fig.3b). With the measurement of the phase information, however, it becomes possible to see the unexpected diagonal line-shape for LSCO (Fig.2f), which reveals the correlation of emission energy and coherence energy over the broad distribution of coherence energies excited.

**Model.** Traditional analysis of MDCS data uses a perturbative approach to solve the Liouville variant of the Schrodinger equation in a single particle framework, with discrete states(*51*). For strongly correlated materials, such as optimally doped LSCO, the many-body nature of the different excitations and states should be taken into account for a full description and understanding of the experimental results. However, as we show below, a single particle picture can still qualitatively reproduce the results and provide some physical insight into the correlations of this material. Here we consider a modified three-level system, where each level is broadened, but the broadening between the top and bottom level is correlated, as shown in Fig.4a. After the first two pulses, the coherent superpositions between bands 1 and 2 (indicated by coloured dashed arrows) are initiated with a distribution of coherence energies. The third pulse drives the system into a coherent superposition between bands 2 and 3 (black arrow), which radiates to give the signal (coloured solid arrows pointing downwards), leaving the system in band 2. This pathway is depicted in the Feynman-Liouville diagram in Fig.4b. This modified three-level system shows how the correlation between the energy of the states in bands 1 and 3 leads to a correlation between the coherence energy, $E_2$, and the emission energy, $E_3$: an increase in coherence energy leads to a decrease in emission energy, and vice versa.

Fig. 4c-f shows the result of simulations in this framework, where we have introduced phenomenological decay terms to account for decoherence and relaxation, and spectral widths to account for the broadening of the low-energy quantum coherence and the correlation between band 1 and 3. Full details of the simulations, which are based on standard approaches for MDCS experiments, are given in Methods. As mentioned earlier, there are at least two factors that can affect the cross-diagonal width of the peak: correlation strength and decoherence. Thus there is not a unique set of parameters for a given 2D spectrum. In the case of the data in Fig.4c-f we used a Lorentzian linewidth of 80 meV for the transition between bands 1 and 3, decoherence time of 500 fs for the coherences between bands 1 and 2, and a width for the effective correlation of 4 meV (see Methods for details). It can be seen that the key experimental features, (i.e. the extended signal in $t_2$ and the narrow diagonal peak in the 2D spectrum) are all reproduced in this simplified model. A corollary of the correlations observed, is that in the 2D time-domain plot the signal is shifted to negative times. This is observed in both experiment and simulation and can be described in the framework of "fast-light"(*52–54*) or pulse shaping, where a spectral phase gradient is generated by the distribution and evolution of coherence energies as described in SI and Fig.S3.

**Signal origin.** The equilibrium band structure of optimally doped LSCO reveals a continuous density of states from well-below the Fermi energy level ($E_F$) to well above it(*55*). It is then possible that states from about 1.55 eV (the laser photon energy) below $E_F$ to ~1.55 eV above $E_F$ can be involved (see SI for possible pathways). Previous pump-probe experiments indicate that high-energy optical excitations decay on the 10's of femtosecond timescale(*16, 23, 27*), while longer lifetimes have been found by ARPES measurements for excitations around $E_F$(*56*). This suggests that the quantum coherent superpositions that persist for ~500 fs should involve states close to the $E_F$. The first two interactions could then be considered as a coherent Raman-like process where we ignore any resonant interactions involving the first two pulses with higher energy excitations. This is justified by data taken as a function of the delay between the first two pulses, $t_1$, and the corresponding 2D spectrum correlating $E_1$ and $E_2$, which shows none of the correlations seen in Fig.2 (see SI Fig.S4).

Previous Raman measurements on similar samples revealed a smooth, essentially featureless spectrum over the range of $E_2$ energies probed here(*7, 11*). Raman modes of different symmetry (primarily $A_{1g}$, $B_{1g}$, and $B_{2g}$) have been identified by polarization controlled experiments, with the two modes with B symmetry ($B_{1g}$ and $B_{2g}$) being of particular interest for their possible role in the pairing mechanism(*30, 57*). We have repeated our MDCS measurements with different polarization combinations, as shown in Fig.5. These results reveal little difference between co-linear and co-circular excitation, but the peak intensity drops to 50% of the amplitude for cross-circular and ~10% for cross-linear excitation. With the caveat that we expect that the surface stress is minimal in our samples(*58*), these results are not consistent with the trends expected for a Raman process involving modes with A or B symmetry: for $B_{1g}$ or $B_{2g}$ symmetry, one would expect no difference between co-linear and cross-linear excitation, but complete cancellation of the signal in going from co-circular to cross-circular(*7*). For acoustic modes with $A_{1g}$ symmetry, one would expect complete cancellation for both cross-linear and cross-circular polarization sequences(*7*). These results confirm that while the signal we observe is similar to a simple Raman measurement with non-resonant optical fields, this cannot completely describe the results, and the interaction with a higher energy level (band 3 in Fig.4) is required. Finally, regardless of the possible surface stress(*58*), the signal is strongly suppressed for the cross-linear configuration, meaning that the excitation triggered by the third pulse should be close to a linear dipole transition, like a charge-transfer excitation.

**Conclusions**

Here we applied MDCS to the prototypical cuprate superconductor LSCO and found novel quantum signatures: a 500 fs long coherent signal originating from the superpositions of many-body states at low energy. We found a strong correlation between the energy of this coherence and the optical energy of the emitted signal, proving there is interplay between the quantum superpositions at low and high energy in these systems. We stress that this kind of quantum signal, characterized by a continuum distribution of coherence energy and a narrow elongation, has, to our knowledge, never been observed before in any other material.

A fundamental concept valid for correlated electronic materials, which is at odds with single-particle theories, is that when the number of available excitations change by e.g. chemical substitution(*5*) or photo-doping(*18*, *19*), so too does the band structure(*6*, *9*, *55*). Thus the effective energy structure of the sample can change transiently as the measurement proceeds, which can lead to links between low-energy excitations and optically excited electronic states. We propose that it is these interactions that lead to the correlations that we measure between the coherent quantum superpositions with energy 20-60 meV and the optically driven states ~1.55 eV above this.

Our results demonstrate that MDCS is a powerful and novel tool to study electronic correlations and opens the door to addressing directly the quantum-coherent state of many-body matter.

**Materials and Methods**

**Sample details**. The crystal was grown by using the flux technique. Stoichiometric amounts of high purity (>99.99%) $La_2O_3$, $SrCO_3$ and $CuO$ chemicals were mixed. Further 75 mole percent of CuO was added as flux. These powders were placed in a platinum crucible and covered tightly with a platinum lid and heated in a chamber furnace. The crucible was heated to 1250 °C and kept there for 10 hr, then it was slowly cooled down to 1100 °C at a rate of 2 °C/hr. After the growth, the flux was separated by inverting the crucible when it was at 1100 °C. Finally the furnace was cooled down to room temperature at the rate of 50 °C/hr(*60*). The grown crystals were annealed under oxygen flow in order to increase the $T_C$ and cooled very slowly down to room temperature, so the strain should be reduced to a minimum. The Sr doping was 0.15 and it was checked by EPMA technique. The magnetic properties were measured using SQUID magnetometry.

**Measurements details**. The transient FWM experiments were performed using the experimental apparatus described previously(*49, 61*). It utilizes a Ti:Sapphire oscillator generating pulses centred at ~1.55 eV and spectrum shown in Fig.1b (black) at a repetition rate of 93 MHz. The laser output is compressed using a folded prism compressor and split into the four beams in a box geometry by a beam shaper, consisting of a 2D grating pattern applied to a spatial light modulator (SLM). The four beams are collected by a lens and propagate through all common optics, in a 4F geometry, to ensure a high level of phase stability between all four beams. They are incident on a SLM-based pulse-shaper, which allows independent control of the spectral amplitude and phase for each of the four beams. The spectral phase is first optimised to ensure properly compressed pulses. The spectral amplitude of the three excitation pulses and local oscillator (LO) are shaped to give the spectra shown in Fig.1b (coloured). To control the timing of each pulse, and hence the inter-pulse delays, independent spectral phase gradients are applied to each pulse. The LO is attenuated and delayed by a neutral density filter and the 4 beams are focussed onto the sample with a 20 cm lens.

The signal generated in the phase matched direction $-(-k_1+k_2+k_3)$ is spatially overlapped with the LO, which co-propagates into a spectrometer where a spectral interferogram is detected, providing details of the signal amplitude and phase. Other signals originating from different wave-mixing are emitted in very different directions and are not detected. An 8-step phase cycling procedure is used to isolate the signal from background scatter and enhance the signal to noise ratio.

In the measurements reported here, data is acquired by setting the delay between the first and second pulses to zero ($t_1=0$) and increasing the delay of the third pulse ($t_2$) by controlling its spectral phase. From this series of interferograms the time resolved signal can be obtained as shown in Fig.2a. Fourier transforms of the data with respect to $t_2$ leads to the 2D spectra in Fig.2d-f. The average power in each excitation beam for these experiments was 400 μW.

Additional measurements at different intensities were performed to ensure these experiments are being done in a third-order regime. The linear polarization dependent experiments were performed by introducing a half-wave plate and linear polarizer into each beam and setting the polarizations accordingly. For the circularly polarized experiments (Fig.5) an additional quarter-wave plate was introduced after the linear polarizers to convert the linear polarizations to circular polarization. The LSCO samples were mounted in a temperature-controlled recirculating cryostat. Measurements were recorded with the sample temperature nominally at 12 K and 66 K and give consistent results indicating the superconducting phase is thermally melted (see SI).

The laser polarization is in the copper-oxygen a-b plane of LSCO, and the signal is independent of the sample orientation in the a-b plane (e.g. if the laser fields are parallel, perpendicular, or in between a and b).

**Simulation**. We simulate the nonlinear response of LSCO by numerically solving the Liouville variant of the Schrodinger equation for the three level system described in Fig.4a,b:

$$i\hbar\dot{\rho} = [H, \rho]$$

with $\qquad H = H_0 + H_{int} + H_{relax}$

$$H_0 = \begin{bmatrix} \epsilon_1 & 0 & 0 \\ 0 & \epsilon_2 & 0 \\ 0 & 0 & \epsilon_3 \end{bmatrix}, \qquad H_{int} = E(\boldsymbol{k}, t)\begin{bmatrix} 0 & 0 & \mu_{13} \\ 0 & 0 & \mu_{23} \\ \mu_{31} & \mu_{32} & 0 \end{bmatrix}; \qquad [H_{relax}, \rho] = i\hbar\gamma_{ij}\rho_{ij}$$

where $\epsilon_i$ is the energy of state $i$, $\mu_{ij}$ is the transition dipole moment for the transition between state $i$ and $j$; $\gamma_{ij}$ is the phenomenological decay rate for each term in the density matrix, $\rho_{ij}$, which corresponds to a population lifetime where $i = j$, and decoherence time where $i \neq j$. $E(\boldsymbol{k}, t)$ is the electric field of the excitation pulses given by

$$E(\mathbf{k}, t) = E_1(t)e^{i(\mathbf{k_1}\cdot\mathbf{r}-\omega_1 t)} + E_2(t)e^{i(\mathbf{k_2}\cdot\mathbf{r}-\omega_2 t)} + E_3(t)e^{i(\mathbf{k_3}\cdot\mathbf{r}-\omega_3 t)}$$

where $\mathbf{k_i}$ is the wave-vector, $\omega_i$ the frequency, and $E_i(t)$ the envelope function of pulse $i$. The electric field is treated perturbatively, and the system of equations expanded up to 3$^{rd}$ order, with the emitted signal being proportional to the third order polarization, corresponding to $\mu_{13}\rho_{13}$, with wavevector $-\mathbf{k_1} + \mathbf{k_2} + \mathbf{k_3}$. The resultant set of interdependent differential equations is solved numerically to facilitate the use of realistic pulses with finite duration. We take the dipole moment for the transition between states 1 and 3 ($\mu_{13}$) to be the same as the one between states 2 and 3 ($\mu_{23}$). The homogeneous linewidths of these transitions ($\gamma_{13}$ and $\gamma_{23}$) were set to 80 meV, such that the ~1.55 eV coherent superpositions decay within the pulse duration, as observed in the projection of the 3D spectrum onto the one-quantum 2D spectrum shown in Supplementary Fig. S4. The homogeneous linewidth $\gamma_{12}$ of the zero-quantum coherence transition was set to 8 meV, corresponding to a decay time of 500 fs. The energy of states 1 and 3 were kept fixed at 0 eV and 1.54 eV and, to account for the distribution of coherence energies, the simulation was run for different values of $\epsilon_2$ and the results summed, weighted by a Gaussian distribution centred at 40 meV with 40 meV FWHM. By varying the energy of state 2, the energy difference between 1 and 2 and between 2 and 3 are anti-correlated, as described in Fig.4a and observed in the experiments. The strength of the correlation between band 1 and 3 is controlled by adjusting the effective spectral width of the third pulse, which drives the transition indicated by the black line in Fig.4a. For the data shown in Fig.4e, corresponding to the case where the long-lived coherences were observed, this width was set to 4 meV. For Fig.4d, where the pulse overlap regime is simulated, this width was set to 30 meV, and in Fig.4c, the sum of these two contributions is plotted.

**Supplementary Materials**

Possible pathways for excitation of the 40±20 meV coherent superpositions

Heating effects due to laser excitation

Emission at negative times

Correlations between $E_1$, $E_2$, and $E_3$

Fig. S1. Signal pathways.

Fig. S2. Thermal effects.

Fig. S3. Apparent signal at negative delays.

Fig. S4. 3D spectra from LSCO and projection onto the 2D planes.

**Acknowledgements**

**General**: FN acknowledges F. Cilento, D. Fausti, and is particularly grateful to A. Avella, for discussions.

**Funding**: JAD acknowledges support from the Australian Research Council.

**Author Contributions**: FN proposed the experiment. JT measured with help from FN. JT and JAD developed the optical setup. JT analysed the data and prepared the figures with help from FN and JAD. JAD developed the model and performed the simulations. DP fabricated and provided high quality samples. JAD and FN wrote the manuscript with help from all the authors.

**Competing interests**: The authors declare no competing financial interests.

**Data and materials availability**: The source data for the figures in the manuscript and Supplementary information and all other data that support the findings of this study are available from the corresponding author on reasonable request.

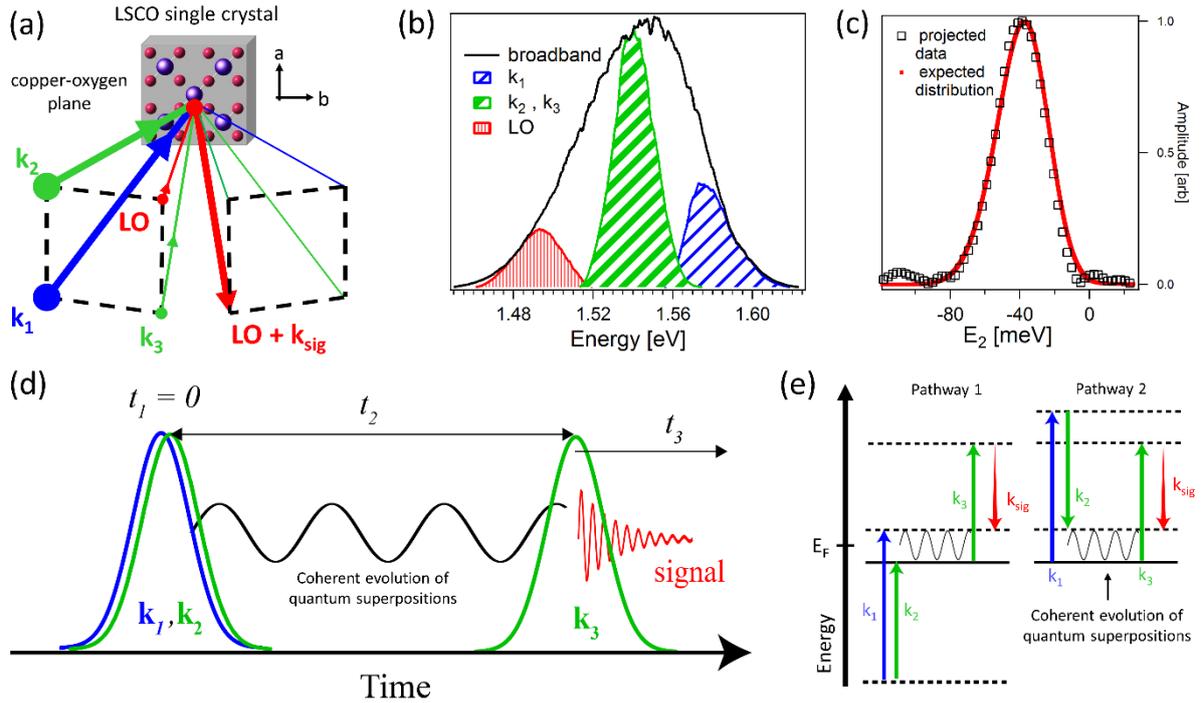

**Fig. 1. Experimental scheme.** (a) The geometry of the excitation is such that the three excitation pulses with wave-vector $k_1$, $k_2$ and $k_3$, form three corners of a box with the local oscillator (LO) on the fourth corner. The excitation is close to normal to the copper-oxygen plane, and the signal which is overlapped with the LO is measured in the reflected direction. (b) The shaped spectrum for each of the excitation pulses and the LO are shown together with the unshaped laser spectrum (black). (c) The convolution of $k_1$ and $k_2$ should determine the range of coherences that can be excited, shown in red. The data from the impulsive response from the sample (black squares) confirms this. (d) The first two, non-degenerate, pulses are overlapped in time, generating quantum superpositions between states separated by energies dependent on the pulse spectra shown in (c). The third pulse arrives at a controllable time, $t_2$, later and probes the coherent evolution of these superpositions through the emission of the signal (red). (e) Two possible signal pathways are depicted for the case of discrete energy levels. The coherent superpositions that are probed can be excited by two "upward" transitions (pathway 1) or an "up" and "down" transition (pathway 2) similar to a coherent Raman-like process.

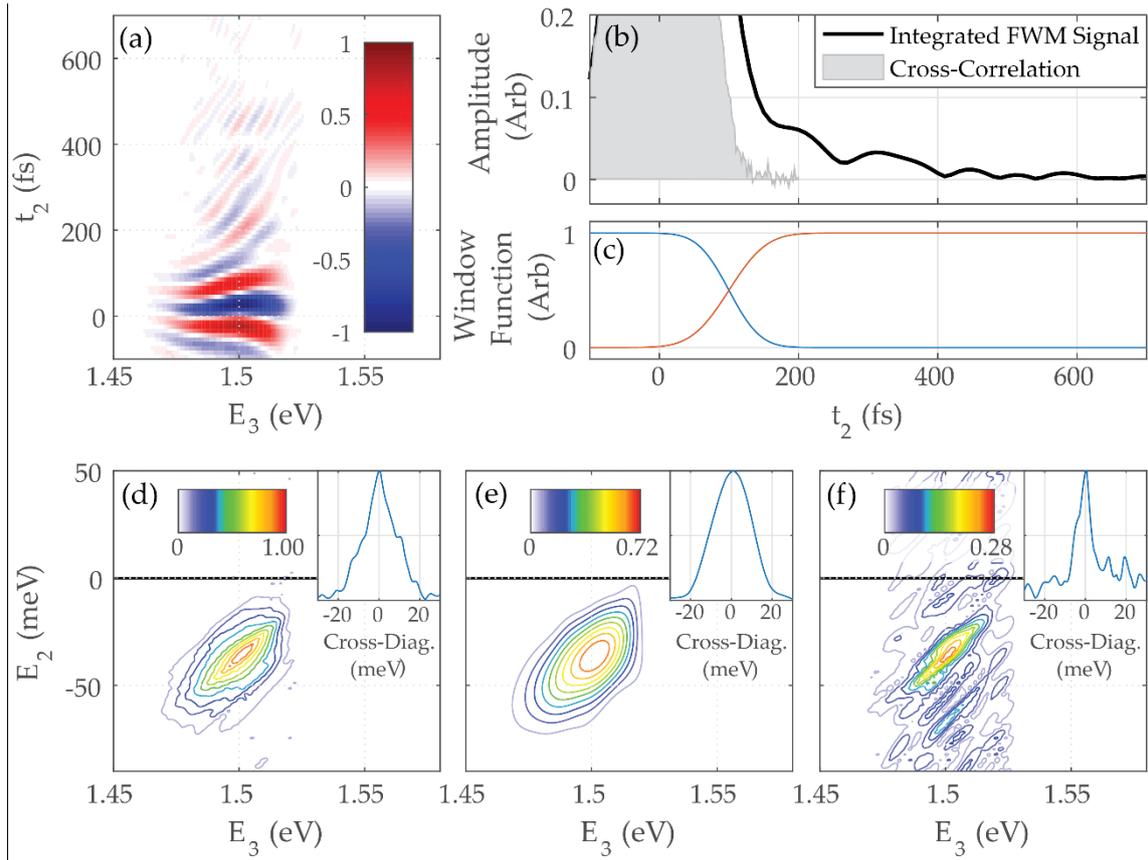

**Fig. 2. Coherent dynamics of the low-energy excitations in LSCO.** (a) The real part of the spectrally resolved signal as a function of $t_2$: the delay between second and third pulses. The oscillating amplitude originates from the phase evolution of the coherent superposition excited by the first two pulses. (b) The spectrally integrated amplitude plotted as a function of $t_2$ shows the decay of the signal extending beyond 500 fs. The pulses cross-correlation is shown in grey. (c) Window functions used to isolate the response during pulse overlap (as wide as the cross-correlation of the pulses) and the response beyond pulse overlap. (d) The Fourier transform of the data with respect to $t_2$ yields the 2D spectrum, where $E_2$ corresponds to the coherence energy i.e. the energy difference between the states in the quantum superposition. By windowing the data using the window functions shown in (c), the response at pulse overlap (e) and after pulse overlap (f) can be isolated. In (f) the diagonal peak shape corresponding to the extended signal becomes clear. In the insets of (d – f) a cross-diagonal slice of the respective 2D spectrum is plotted.

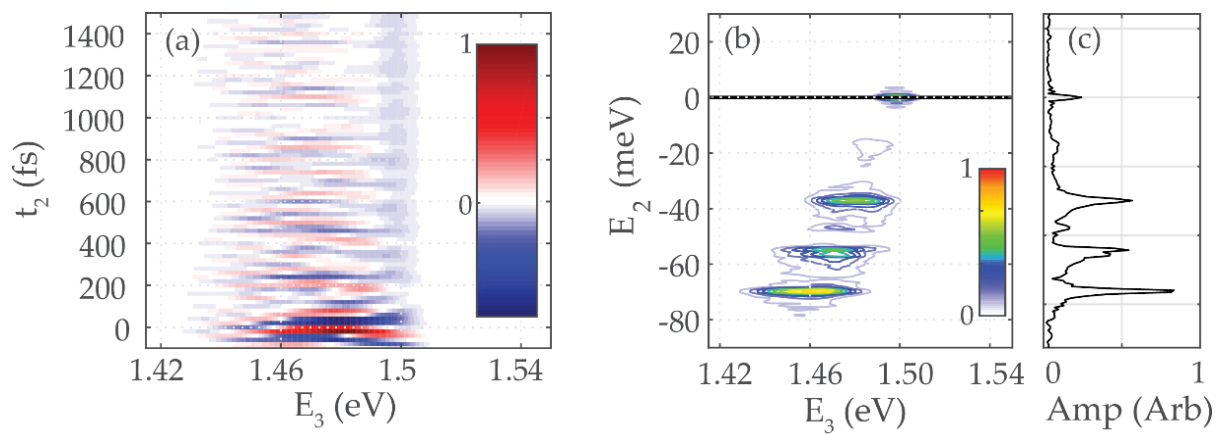

**Fig. 3. Comparison with vibrational coherences.** (a) The real part of the signal from IR812 laser dye, as a function of $t_2$ and $E_3$. The response extends beyond 1500 fs with several oscillation frequencies. The normalized 2D spectrum, (b), and projection on to the $E_2$ axis, (c), shows discrete narrow peaks corresponding to the discrete vibrational modes(*59*). Here each peak is broad along the $E_3$ axis, matching the width of the third pulse. The signal amplitude beyond pulse overlap is about 10x larger for the molecular system (b) than for LSCO (Fig.2f).

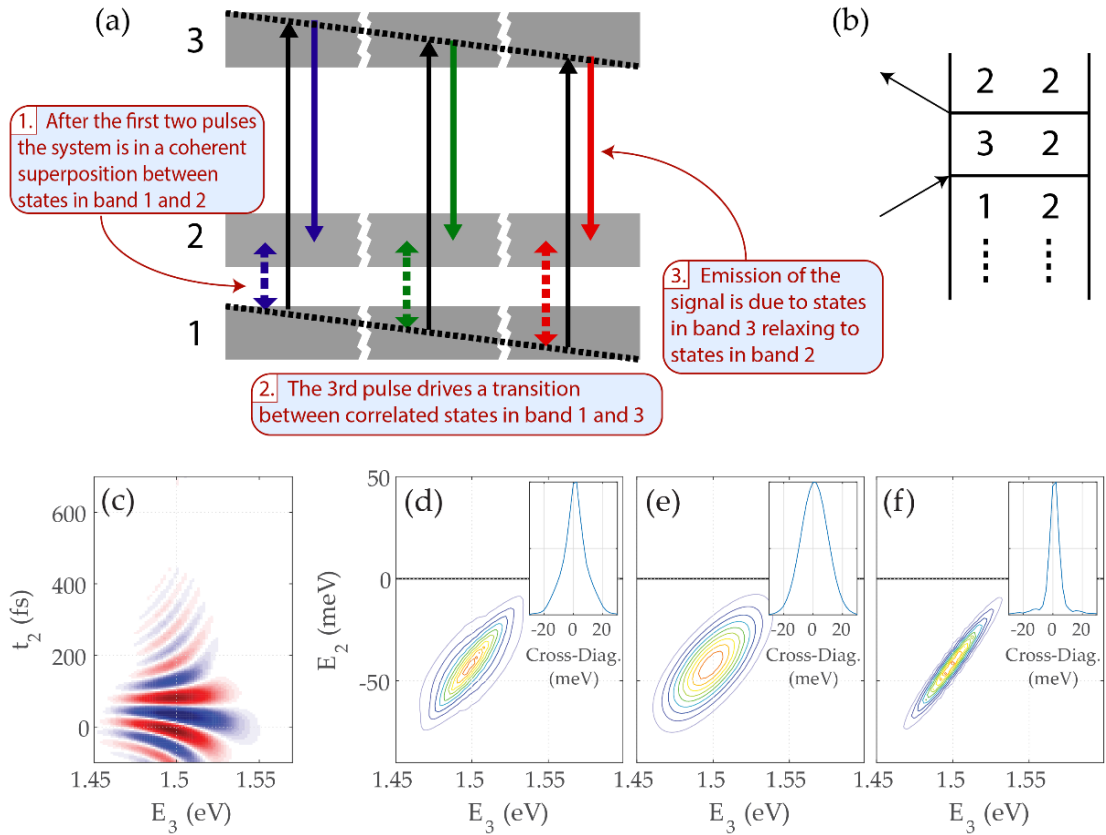

**Fig. 4. Model and simulations for coherent response in LSCO**. (a) The simple single-particle model shows the first two pulses generating quantum superpositions between bands 1 and 2 (dashed coloured arrows), with a distribution of coherence energies. The third pulse (solid black vertical arrows) drives the transition to band 3 with energies correlated to those in band 1 (black dotted lines), leading to emission (solid coloured arrows) which is anti-correlated with the coherence energy. (b) The corresponding Feynman diagram beginning with the coherent superposition between states in bands 1 and 2 triggered by the first two light-matter interactions. In panels (c-f) we show the results of simulations based on this picture. (c) The evolution of the real part of the signal, (d) the 2D spectrum of the full response, (e) the 2D spectrum from the response during pulse and (f) the 2D spectrum from contribution making up the extended signal without pulse overlap. The insets show the corresponding cross-diagonal slices.

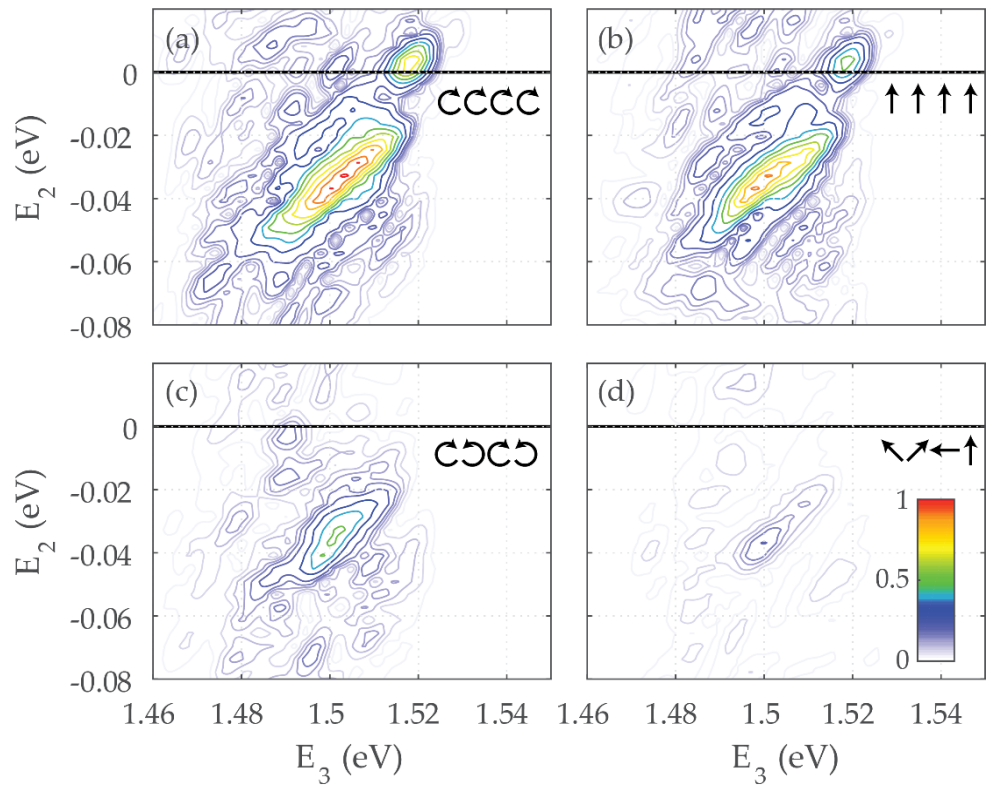

**Fig. 5. Polarization dependent 2D spectra in LSCO**. The 2D spectra from the extended part of the signal for (a) co-circular, (b) co-linear, (c) cross-circular and (d) cross-linear excitation are shown. The narrow diagonal peak shape remains unchanged, only the amplitude varies for the different polarization combinations.